# From Good Practices to Effective Policies for Preventing Errors in Spreadsheets

Daniel Kulesz
Institute of Software Technology, University of Stuttgart, Germany
daniel.kulesz@informatik.uni-stuttgart.de

**ABSTRACT**

*Thanks to the enormous flexibility they provide, spreadsheets are considered a priceless blessing by many end-users. Many spreadsheets, however, contain errors which can lead to severe consequences in some cases. To manage these risks, quality managers in companies are often asked to develop appropriate policies for preventing spreadsheet errors.*
*Good policies should specify rules which are based on „known-good" practices. While there are many proposals for such practices in literature written by practitioners and researchers, they are often not consistent with each other. Therefore no general agreement has been reached yet and no science-based „golden rules" have been published.*
*This paper proposes an expert-based, retrospective approach to the identification of good practices for spreadsheets. It is based on an evaluation loop that cross-validates the findings of human domain experts against rules implemented in a semi-automated spreadsheet workbench, taking into account the context in which the spreadsheets are used.*

## 1 INTRODUCTION

Since their invention in the late 1970s, spreadsheet software packages have been valued as a priceless blessing by millions of end-users. Thanks to the huge flexibility they provide, end-users can shape their own computing solutions even for complex computing tasks. This does not only come in handy for individuals and small businesses who do not have the budget for ready-made solutions. Spreadsheets are used on a large scale in enterprises as well – especially in situations where the IT department fails to provide acceptable solutions on time or budget.

There is a steadily growing body of evidence which suggests that many of the spreadsheets in use today suffer from a high percentage of errors [Panko, 2008] [Powell et al., 2007]. Several horror stories from the past like [Godfrey, 1995] indicate that errors in spreadsheets can lead to wrong results which, in turn, can result in costly wrong decisions. Although many executives and senior managers rate the impact of spreadsheet errors as less critical [Caulkins et al., 2008], several governments regard them as a serious threat and have already reacted by issuing laws like Sarbanes Oaxley Act 404, Basel II or Solvency II which try to confine the uncontrolled use of spreadsheets.

Apart from the fact that laws can only address the spreadsheet quality problem in limited scope, there is no generally accepted recommendation among researchers either. After a critical review of the literature on spreadsheet errors, Powell, Baker and Lawson [Powell et al., 2008] draw several alarming conclusions about the current „state of the art":

- There is no generally accepted way of counting and classifying errors in spreadsheets. Although a few taxonomies exist (i.e. [Panko and Halverson, 1996] [Rajalingham et al., 2000]), all of them are not clearly defined, untested and their applicability is limited to a few, implicitly defined contexts. (Note: Some of this criticism was meanwhile addressed in [Panko and Aurigemma, 2010])

- The real quantitative (economic) impact resulting from spreadsheet errors is largely unknown.

- Existing studies about the frequency of spreadsheet errors are not comparable due to a lack of standardization.

- There is very little insight about where spreadsheet errors originate from and how to prevent them. Existing studies have been suggestive and based mostly on observations in laboratory experiments which cannot be applied to operational spreadsheets in the field.

- Approaches which claim to be effective in the detection of errors in spreadsheets are not described with enough details. Furthermore, it is mostly unknown how they compare in terms of effectiveness and efficiency.

Despite these significant uncertainties in spreadsheet error research, the number of commercial offerings for addressing spreadsheet quality problems is increasing steadily. The solutions offered range from error-detection tools, end-user training, consultancy for company policies, refactoring or reengineering of existing spreadsheets, to full migration projects towards „professional" software. There is also a large amount of literature available where experienced practitioners proclaim „good" and „poor" practices regarding development, exploitation and management of spreadsheets. Some prominent examples include [Read and Batson, 1999], [Raffensberger, 2008], [O'Beirne, 2005], [Bovey et. al., 2009] or [Powell and Baker, 2010]. Particularly remarkable about these recommendations is that their authors have different opinions about some of the discussed practices.

This vast gap between practice and science leads to endless debates about basic principles for the proper handling of spreadsheets on various abstraction levels. For instance, on the process level there is a basic discussion about the right mixture between issuing regulatory steps and preserving the end-users' freedom. On the implementation level, basic questions such as whether to break down a spreadsheet into separate areas, to use named ranges or to put multiple values in a single cell tend to split parties into separate camps.

At first glance, it seems obvious that the spreadsheet community must proceed towards an agreement on Best Practices for spreadsheets. But it is not clear whether this demand can be satisfied at all: Colver [Colver, 2004] and Grossmann [Grossman, 2002] argue that spreadsheet practices are always context-dependent, while Dunn [Dunn, 2010] claims that there must exist a large body of universal practices. At least, there seems to be a general agreement that certain practices which can be recommended in certain contexts do exist.

The current situation is not satisfactory, especially for quality managers who are expected to issue policies for the use of spreadsheets in their companies. They have a sound claim for science-based practices which reduce both the frequency and the impact of errors in their spreadsheets. We need to find a way to measure which practices lead to which kind of effects on spreadsheet quality when applied in which contexts. After identifying such practices, we should be able to gain more insight into the question whether good practices exist which can be applied independently of the spreadsheet's context.

This paper proposes an expert-based, retrospective approach to the identification of good practices for spreadsheets. Since the approach relies on a tool which is still under development, we cannot provide any concrete results yet. Nevertheless, we would like to present some details about our ongoing work and compare its conceptual direction with other approaches.

## 2 APPROACH

Our approach is based on a retrospective analysis of typical, existing spreadsheets – not on a greenfield strategy. There are numerous promising „Best Practice" approaches to building better spreadsheets from scratch by using fundamentally different development techniques like [Dunn, 2009], [Paine, 2001] or [Grossman and Özlük, 2010] or switching to alternative technologies like [Miller, 2010]. We have chosen not to follow this path for several reasons:

- Time is a critical factor. There are incredible amounts of poor spreadsheets in use *today*. Of course, rewriting them from scratch might pay off somewhere in the future. But you won't get it done in the *near* future. We need a cheap yet effective short-term solution for the immediate problems.

- Convincing users is hard because users are conservative. Even if you succeed in developing a better spreadsheet approach, you still have to convince the end-users of its usefulness. This is rather unlikely to happen: Several conceptually promising approaches to re-defining the nature of a spreadsheet like Lotus Improv or its successor Quantrix Modeler never managed to reach and convince such masses as VisiCalc's logical successors Lotus 1-2-3 and Microsoft Excel did.

- Experience is an asset, and learning new paradigms is hard. Even if you manage to convince the end-users of a new paradigm's benefits, they will still have to learn to cope with it in practice. Their experience with existing tools and techniques could become worthless if the new paradigm is too different from what they know and how they use to work.

Therefore, instead of searching for new practices to (re)create better spreadsheets, we are trying to find out which of the already known practices are favourable. Because many of these practices have penetrated some of today's spreadsheets, we are attempting to investigate these spreadsheets, identify the practices used there and review their effectiveness based on the quality of the spreadsheets.

Figure 1 illustrates our approach. Briefly, it works as follows: First, we specify a policy consisting of a set of rules based on suggested practices for spreadsheets from literature. Then we ask a human domain expert to review a given spreadsheet and to provide us with a subjective, general impression about it. In case the domain expert finds any errors, we ask him to report them, too. In parallel, we feed the spreadsheet into our workbench. The workbench checks the spreadsheet for any violations of the previously provided policy and summarizes them in a report. The report includes detailed information about which rules were violated. We repeat the process with other spreadsheets from the same domain, maintaining the same policy.

Finally, we compare the domain expert's impressions with the rule violations reported by the workbench. This allows us to identify rules which correlate with the ratings of the expert. A „perfect" practice for the used context would be represented by a rule which does not report violations for spreadsheets the expert rated as „good" but does report violations for spreadsheets the expert rated as „poor", without producing false positives or false negatives.

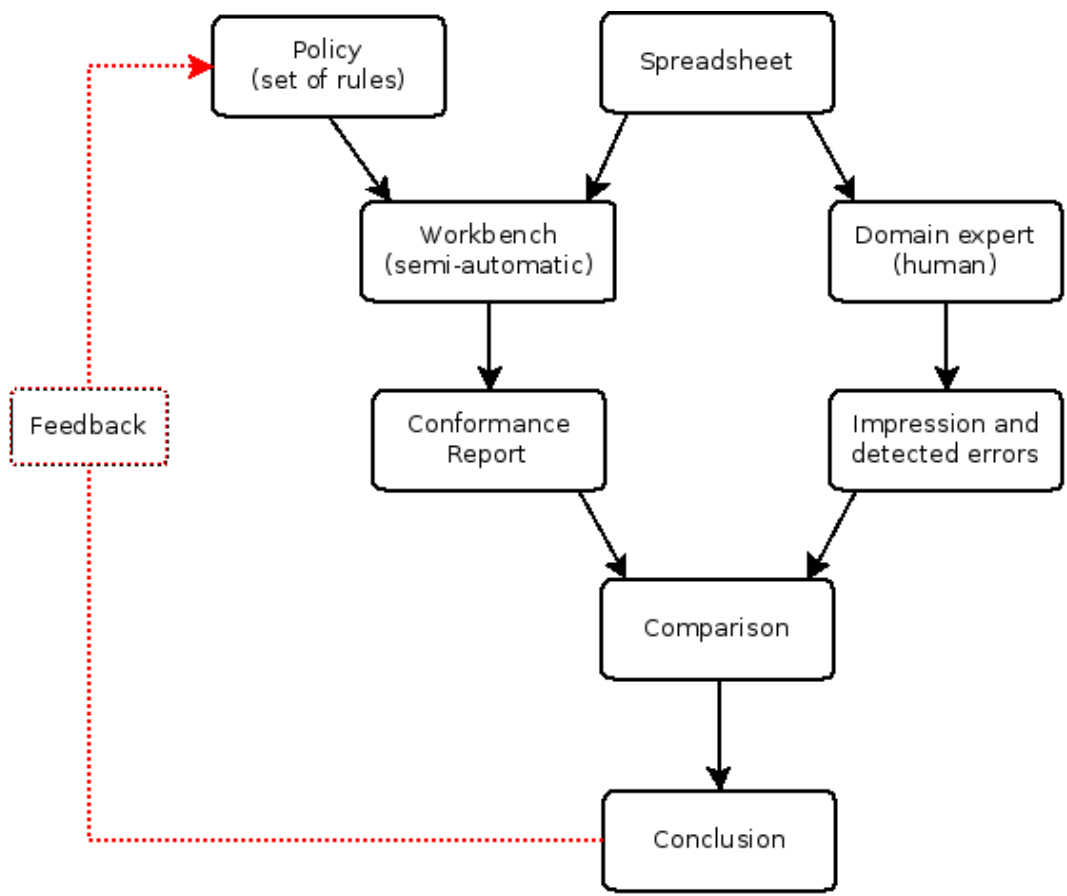

*Figure 1: Our concept for evaluating spreadsheet practices*

## 2.1 Applicability

It is the nature of our retrospective analysis that it is based on the *current state* of the inspected spreadsheet. The approach takes into account the fact that in many cases the steps suggested to be followed by spreadsheet practices leave visible traces in the final outcome. This manifestation makes it possible to verify whether the practices have been followed later on. For instance, it is possible to inspect all formulae in each row and column of a spreadsheet to check whether the practice "Use one formula per row or column" [Read and Batson, 1999] has been followed. In almost the same manner we could check dependencies between cells by analyzing their formulae to see whether the practice "Refer to the left and above" [Read and Batson, 1999] was obeyed.

Our approach is not limited to a specific taxonomy for the classification of spreadsheet errors or quality problems. We leave this choice completely up to the human domain expert who is needed for the manual inspection step in most cases. Theoretically, it would be possible to skip this step provided we had a comprehensive error log for a „known bad" spreadsheet beforehand. Unfortunately, such error logs are very unlikely to be found in spreadsheet samples from the field.

It also does not matter what sort of spreadsheet is to be analyzed and how complex it is – as long as the domain expert is able to understand it, it's fine. But we demand that the domain expert must not be the author of the inspected spreadsheet as spreadsheet users are often over-confident about their own work [Brown and Gould, 1987] [Panko, 2003]. It is obvious that different experts might have different perceptions of „good" and „poor" spreadsheets. Our underlying assumption is that if many experts rate a specific spreadsheet as „poor", there must be something wrong with it.

**2.2 Limitations**

Currently, our approach does not investigate *prior states* of the inspected spreadsheet. Thus, we cannot derive any conclusions about the *process* it has undergone. This includes the spreadsheet's initial creation as well as any manipulations performed on it until it reached the current state. As a consequence, the approach does not allow the evaluation of process-oriented development practices like „Write your application in the earliest version of Excel that you expect it to run in" [Bovey et. al., 2009] or „Create and run test cases covering all logic paths" [O'Beirne, 2005]. Without analyzing prior versions of spreadsheets we also cannot distinguish between „bad" spreadsheets and „good spreadsheet that go bad during usage" [Baxter, 2010].

Another limitation is that our approach relies on a tool for the automated checking of spreadsheets against previously specified rules. Therefore, we can only evaluate spreadsheet practices that can be checked by computer programs.

**3 SPREADSHEET WORKBENCH**

The development of our spreadsheet workbench aims at offering a multi-purpose „Swiss-Army-Knife" tool for the quality assurance of spreadsheets. Unlike other tools which are mostly implemented as commercial add-ins for existing spreadsheet programs, our workbench is a stand-alone, cross-platform tool written in Java and targeted to a broad audience:

- *End-users* can find quality problems in their spreadsheets before sharing their spreadsheets with peers or basing decisions on them.

- *Quality Managers* can define organizational policies for spreadsheets based on concrete practices accepted by their organization.

- *Spreadsheet auditors* can check whether both external and the internal end-users' spreadsheets comply with the organizational policies.

- *Researchers* can use the workbench as a tool to verify or falsify hypotheses about effectiveness and efficiency of spreadsheet practices.

Providing a detailed description of all the workbench's features would exceed the scope of this paper. Instead, we want to provide an example which illustrates how researchers following a retrospective analytical approach (like, but not limited to, an approach as described above) could use the workbench.

**3.1 Creating a new policy**

Policies are represented in the workbench by so-called „scenarios" while spreadsheet practices are implicitly defined by so-called „rule checkers". For instance, there could be a rule checker which checks for the presence of hardcoded constants in formulae.

A scenario defines a set of concrete rule checkers to be used. The rule checkers themselves also have customizable parameters, so a scenario represents a catalogue of individually configured rule checkers Our current UI prototype for this customization is shown in Figure 2.

In the initial version of the spreadsheet workbench, end-users are not able to specify new rule checkers themselves. But unlike tools such as Spreadsheet Professional [Spreadsheet Innovations, 2011] or XL Analyst [Codematic, 2011] the rules are not hardcoded. Instead, the workbench provides a Java API which allows professional programmers to implement their own rule checkers as plug-ins. For later versions of the workbench it is planned to replace this mechanism by a more end-user friendly facility.

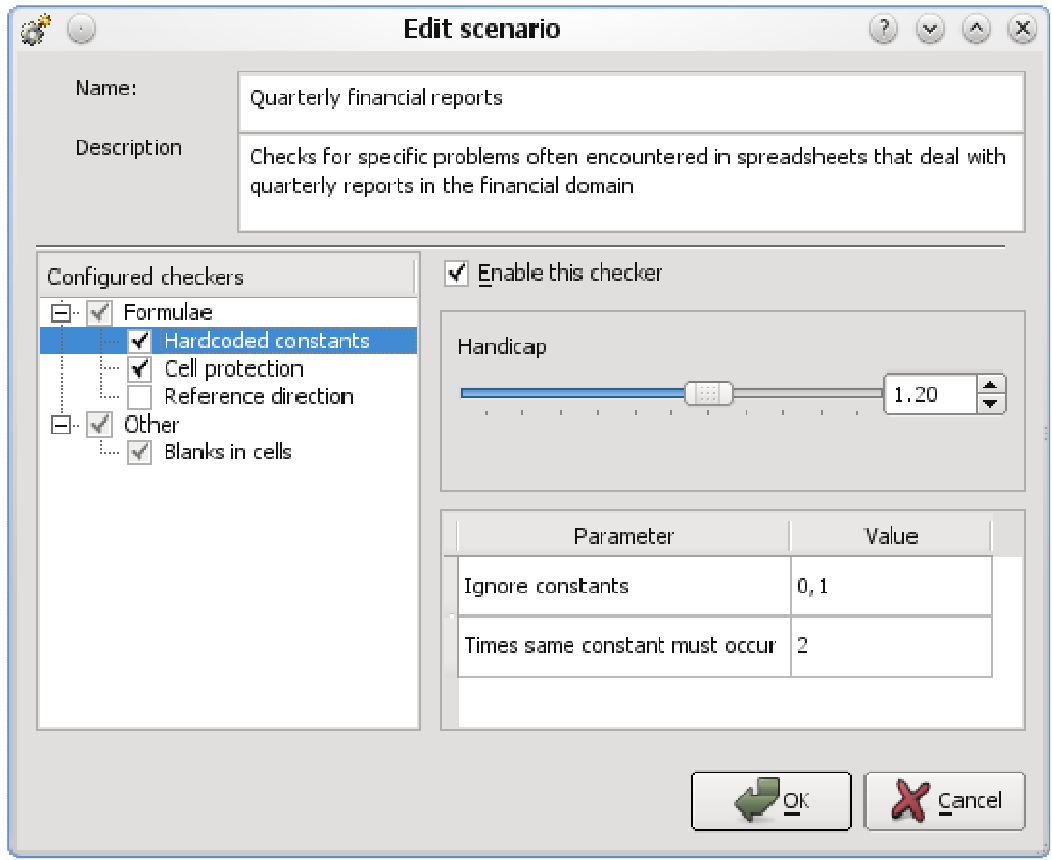

*Figure 2: UI Prototype of the spreadsheet workbench's rule checker customization dialog*

We assume that a professional programmer has already implemented four rule checkers, which check for the following criteria:

- Are there hardcoded constants which are used in multiple formulae?

- Are there formulae in cells which do not have cell protection enabled?

- Are there formulae which refer to the right or below?

- Are there cells which consist only of one or more blanks (spaces)?

The first step for creating a new scenario is to select the desired rule checkers. To simplify matters we assume that we want to use the first two of the above rules for a scenario named „quarterly financial reports".

Each rule checker comes with a default configuration for its customizable parameters. For most rule checkers, their parameters allow to define thresholds and exceptions.

For instance, the „constants in formulae" checker could have the option to ignore certain constants. This could come in handy if the spreadsheets in a scenario deal with percentages and thus often contain the constant „1". Some practitioners argue that it is easier to read and verify formulae which contain values and not cell references in formulae, if the affected value is short, simple and not expected to change (i.e. because it is a natural constant).

**3.2 Running the analysis**

After finishing the customization of the scenario, the workbench can be instructed to „run" the scenario on a set of given input spreadsheets. Internally, the workbench makes use of an abstract spreadsheet model which allows it to bind to various spreadsheet formats and locations using external APIs. For the initial version, it is planned to support Microsoft Excel workbooks, OpenOffice.org Calc documents and online spreadsheets from GoogleDocs.

**3.3 Reviewing the results**

Once the workbench is done with the analysis of the input spreadsheets, it summarizes its findings in a report. Figure 3 illustrates our current prototype for this „report view". The findings presented in the report can be filtered and grouped by various criteria. Currently this includes the cells which are affected, the rule checker which reported the findings and the affected spreadsheet.

The workbench also provides a detailed finding view. Using the API offered by the workbench, rule checkers can provide a textual description of the finding, a general explanation why this could lead to problems and a suggestion how the problem could be remedied by using alternative constructs. The workbench's API also allows each rule checker to render own visualizations for its findings.

**3.4 Refining the scenario**

After reviewing the workbench's report, researchers can further refine a scenario by adding or removing rule checkers, tweaking their configurations and re-running the analysis. After several iterations it should be possible to end up with a scenario which is mostly consistent with the issues which the domain expert identified in the manual review. In case this scenario even complies with ratings by other experts from the same domain, there is a chance that we have identified a promising candidate for a domain-specific policy. Shall the scenario even lead to findings which are consistent with the manual review of experts from other domains, then the practices represented by the scenario's rules are probably good candidates for „universally" good spreadsheet practices.

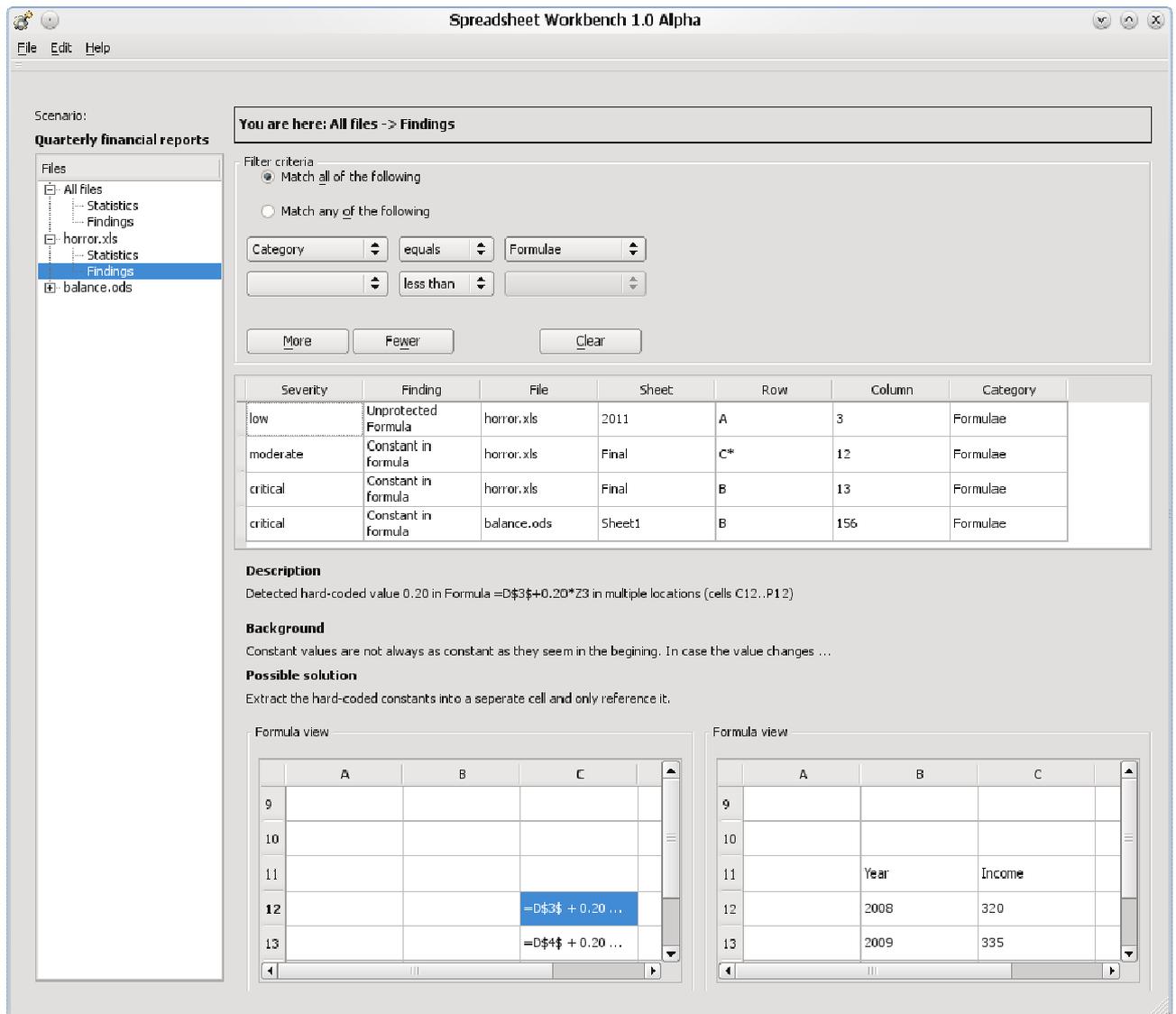

*Figure 3: UI Prototype of the spreadsheet workbench's report view*

**4 CONCLUSION AND FUTURE WORK**

We have proposed an approach towards the identification of good practices for spreadsheets. It is based on an evaluation loop that cross-validates the findings of human domain experts against rules implemented in a semi-automated spreadsheet workbench. Due to its retrospective nature, the approach cannot analyze process-oriented practices, though. For some process-oriented practices we could partly overcome this limitation by extending the analysis on prior versions of spreadsheets as well.

The workbench's development is still underway, so we cannot provide any evidence that the approach will work as expected. But given the case it will, there is a chance for getting closer to the goal of offering a science-based catalogue of accepted good practices for spreadsheets. This would be a vast improvement over the current situation of having to choose from conflicting practitioner recommendations.

## 5 ACKNOWLEDGEMENTS

The author would like to thank Michael Starzmann who contributed greatly to the development of the spreadsheet workbench described in this paper. The author is also grateful to Jochen Ludewig and Kornelia Kuhle for their constructive criticism of earlier versions of this paper.

## 6 REFERENCES

Baxter, R. (2010): "Spreadsheets and Access Databases – Enterprise Control, Efficiency and Insight", Proceedings of the EuSpRIG 2010 conference

Bovey, R., Wallentin, D., Bullen, S., Green, J. (2009): "Professional Excel Development", 2nd edition, Addison-Wesley

Brown, P. and Gould, D.(1987): "An Experimental Study of People Creating Spreadsheets", ACM Transactions on Office Information Systems, Vol. 5, Issue 3

Caulkins, J, Morrison, E., Weidemann, T. (2008): "Do Spreadsheet Errors Lead to Bad Decisions: Perspectives of Executives and Senior Managers", Heinz Research

Colver, D (2004): "Spreadsheet good practice: is there any such thing?", Proceedings of the EuSpRIG 2004 conference

Dunn, A. (2009): "Automated Spreadsheet Development", Proceedings of the EuSpRIG 2009 conference

Dunn, A. (2010): "Spreadsheets – the Good, the Bad and the Downright Ugly", Proceedings of the EuSpRIG 2010 conference

Grossman, T. (2002): "Spreadsheet Engineering – A Research Framework", Proceedings of the EuSpRIG 2002 conference

Grossman, T. and Özlük, Ö. (2010): "Spreadsheets Grow Up: Three Spreadsheet Engineering Methodologies for Large Financial Planning Models", Proceedings of the EuSpRIG 2010 conference

Godfrey, K. (1995): "Computing error at Fidelity's Magellan fund", The Risk Digest, Vol 16, Issue 72, ACM Committee on Computers and Public Policy

Miller, D. (2010): "Sumwise: A Smarter Spreadsheet", Proceedings of the EuSpRIG 2010 conference

O'Beirne, P (2005): "Spreadsheet Check and Control", Systems Publishing

Paine, J. (2001): "Ensuring Spreadsheet Integrity with Model Master", Proceedings of the EuSpRIG 2001 conference

Panko, R., Halverson P. (1996): "A Framework for Research on Spreadsheet Risks", Proceedings of the Twenty-Ninth Hawaii International Conference on System Sciences

Panko, R. (2003): "Reducing Overconfidence in Spreadsheet Development", Proceedings of the EuSpRIG 2003 conference

Panko, R. (2008): "What we know about Spreadsheet Errors", http://panko.shidler.hawaii.edu/ssr/Mypapers/whatknow.htm 7:45pm 03/31/11

Panko, R. And Aurigemma S. (2010): "Revising the Panko-Halverson Taxonomy of spreadsheet errors", Decision Support Systems, Vol. 49, Issue 2

Powell, S., Lawson, B., Baker, K. (2007): "Impact of Errors in Operational Spreadsheets", Proceedings of the EuSpRIG 2007 conference

Powell, S., Baker K., Lawson, B. (2008): "A critical review of the literature on spreadsheet errors", Decision Support Systems, Vol. 46, Issue 1

Powell, S and Baker, K. (2010): "Management Science: The Art of Modeling with Spreadsheets", 3$^{rd}$ edition, John Wiley and Sons

Raffensberger, J. (2008): "The Art of the Spreadsheet", http://john.raffensperger.org/ArtOfTheSpreadsheet, 2:00pm 05/25/11

Rajalingham K., Chadwick D., Knight B. (2000): "Classification of Spreadsheet Errors", Proceedings of the EuSpRIG 2000 conference

Read, N and Batson J. (1999): "Spreadsheet modelling best practice", Accountants Digest, Vol 406, ICAEW

Spreadsheet Innovations (2011): "Spreadsheet Professional – Product Description", www.spreadsheetinnovations.com/Prod_description.htm 3:25pm 03/30/11

Codematic (2011): "XLAnalyst – Proactive Spreadsheet Risk Management", www.spreadsheethell.com/xlanalyst 3:30pm 03/30/11